\documentclass[acmsmall, screen]{acmart}

\makeatletter
\@ifundefined{lhs2tex.lhs2tex.sty.read}%
  {\@namedef{lhs2tex.lhs2tex.sty.read}{}%
   \newcommand\SkipToFmtEnd{}%
   \newcommand\EndFmtInput{}%
   \long\def\SkipToFmtEnd#1\EndFmtInput{}%
  }\SkipToFmtEnd

\newcommand\ReadOnlyOnce[1]{\@ifundefined{#1}{\@namedef{#1}{}}\SkipToFmtEnd}
\usepackage{amstext}
\usepackage{amssymb}
\usepackage{stmaryrd}
\DeclareFontFamily{OT1}{cmtex}{}
\DeclareFontShape{OT1}{cmtex}{m}{n}
  {<5><6><7><8>cmtex8
   <9>cmtex9
   <10><10.95><12><14.4><17.28><20.74><24.88>cmtex10}{}
\DeclareFontShape{OT1}{cmtex}{m}{it}
  {<-> ssub * cmtt/m/it}{}

\DeclareFontShape{OT1}{cmtt}{bx}{n}
  {<5><6><7><8>cmtt8
   <9>cmbtt9
   <10><10.95><12><14.4><17.28><20.74><24.88>cmbtt10}{}
\DeclareFontShape{OT1}{cmtex}{bx}{n}
  {<-> ssub * cmtt/bx/n}{}
\newcommand{\Conid}[1]{\mathit{#1}}
\newcommand{\Varid}[1]{\mathit{#1}}
\newcommand{\anonymous}{\kern0.06em \vbox{\hrule\@width.5em}}

\usepackage{polytable}

\@ifundefined{mathindent}%
  {\newdimen\mathindent\mathindent\leftmargini}%
  {}%

\def\resethooks{%
  \global\let\SaveRestoreHook\empty
  \global\let\ColumnHook\empty}
\newcommand*{\savecolumns}[1][default]%
  {\g@addto@macro\SaveRestoreHook{\savecolumns[#1]}}
\newcommand*{\restorecolumns}[1][default]%
  {\g@addto@macro\SaveRestoreHook{\restorecolumns[#1]}}
\newcommand*{\aligncolumn}[2]%
  {\g@addto@macro\ColumnHook{\column{#1}{#2}}}

\resethooks

\newcommand{\onelinecommentchars}{\quad-{}- }
\newcommand{\commentbeginchars}{\enskip\{-}
\newcommand{\commentendchars}{-\}\enskip}

\newcommand{\visiblecomments}{%
  \let\onelinecomment=\onelinecommentchars
  \let\commentbegin=\commentbeginchars
  \let\commentend=\commentendchars}

\newcommand{\invisiblecomments}{%
  \let\onelinecomment=\empty
  \let\commentbegin=\empty
  \let\commentend=\empty}

\visiblecomments

\newlength{\blanklineskip}
\newcommand{\hsindent}[1]{\quad}% default is fixed indentation
\let\hspre\empty
\let\hspost\empty

\EndFmtInput
\makeatother
\ReadOnlyOnce{polycode.fmt}%
\makeatletter

\newcommand{\hsnewpar}[1]%
  {{\parskip=0pt\parindent=0pt\par\vskip #1\noindent}}

\newcommand{\hscodestyle}{}

\newcommand{\sethscode}[1]%
  {\expandafter\let\expandafter\hscode\csname #1\endcsname
   \expandafter\let\expandafter\endhscode\csname end#1\endcsname}

  {\par\noindent
   \advance\leftskip\mathindent
   \hscodestyle
   \let\\=\@normalcr
   \let\hspre\(\let\hspost\)%
   \pboxed}%
  {\endpboxed\)%
   \par\noindent
   \ignorespacesafterend}

  {\hsnewpar\abovedisplayskip
   \advance\leftskip\mathindent
   \hscodestyle
   \let\hspre\(\let\hspost\)%
   \pboxed}%
  {\endpboxed%
   \hsnewpar\belowdisplayskip
   \ignorespacesafterend}

  {\hsnewpar\abovedisplayskip
   \advance\leftskip\mathindent
   \hscodestyle
   \let\\=\@normalcr
   \(\pboxed}%
  {\endpboxed\)%
   \hsnewpar\belowdisplayskip
   \ignorespacesafterend}

\newcommand{\plainhs}{\sethscode{plainhscode}}

\plainhs

  {\hsnewpar\abovedisplayskip
   \advance\leftskip\mathindent
   \hscodestyle
   \let\\=\@normalcr
   \(\parray}%
  {\endparray\)%
   \hsnewpar\belowdisplayskip
   \ignorespacesafterend}

  {\parray}{\endparray}

  {\(\parray}{\endparray\)}

\def\codeframewidth{\arrayrulewidth}
\RequirePackage{calc}

  {\parskip=\abovedisplayskip\par\noindent
   \hscodestyle
   \arrayrulewidth=\codeframewidth
   \tabular{@{}|p{\linewidth-2\arraycolsep-2\arrayrulewidth-2pt}|@{}}%
   \hline\framedhslinecorrect\\{-1.5ex}%
   \let\endoflinesave=\\
   \let\\=\@normalcr
   \(\pboxed}%
  {\endpboxed\)%
   \framedhslinecorrect\endoflinesave{.5ex}\hline
   \endtabular
   \parskip=\belowdisplayskip\par\noindent
   \ignorespacesafterend}

\newcommand{\framedhslinecorrect}[2]%
  {#1[#2]}

  {\(\def\column##1##2{}%
   \let\>\undefined\let\<\undefined\let\\\undefined
   \newcommand\>[1][]{}\newcommand\<[1][]{}\newcommand\\[1][]{}%
   \def\fromto##1##2##3{##3}%
   }{\) }%

  {\let\orighscode=\hscode
   \let\origendhscode=\endhscode
   \def\endhscode{\def\hscode{\endgroup\def\@currenvir{hscode}\\}\begingroup}
   \orighscode\def\hscode{\endgroup\def\@currenvir{hscode}}}%
  {\origendhscode
   \global\let\hscode=\orighscode
   \global\let\endhscode=\origendhscode}%

\makeatother
\EndFmtInput
\usepackage{iftex}
\ifPDFTeX
\usepackage[utf8]{inputenc}
\fi
\usepackage{verbatim}
\usepackage{verbatimbox}
\usepackage{hyperref}
\usepackage{mathpartir}
\usepackage{algpseudocode}
\usepackage{amsmath}
\usepackage{framed}
\usepackage{graphicx}
\usepackage{caption}

\ifPDFTeX
\usepackage[T1]{fontenc}
\usepackage{zlmtt}
\fi
\usepackage{fancyvrb}
\AtBeginDocument{%
  \providecommand\BibTeX{{%
    \normalfont B\kern-0.5em{\scshape i\kern-0.25em b}\kern-0.8em\TeX}}}

\setcopyright{acmcopyright}
\copyrightyear{2024}
\acmYear{2024}
\acmDOI{10.1145/1122445.1122456}

\acmJournal{JACM}
\acmVolume{1}
\acmNumber{1}
\acmArticle{1}
\acmMonth{9}
\usepackage{listings} %插入代码

\lstnewenvironment{code}{
	\lstset{
		language={},
		backgroundcolor = \color[RGB]{245,245,245},
		basicstyle=\ttfamily\small,
		keepspaces=true,
		numbers=left, %设置行号位置
		numberstyle=\tiny, %设置行号大小
		keywordstyle=\color{blue}, %设置关键字颜色
		commentstyle=\color[RGB]{92,92,92}, %设置注释颜色
		frame= bt, %single, %设置边框格式 bottom and top
		breaklines=true, %自动折行
		extendedchars=false, %解决代码跨页时，章节标题，页眉等汉字不显示的问题
		xleftmargin=0.2em,xrightmargin=-0em, aboveskip=1em, %设置边距
		rulecolor=\color[RGB]{220,220,220},
		showstringspaces=false, %不显示字符串中的空格
		tabsize=3, %设置tab空格数
		stringstyle=\color[RGB]{128,0,0},        %\slshape\color[RGB]{128,0,0},
		showspaces=false, %不显示空格
		escapeinside=``
	}
}{}
\begin{document}

\title{A Top-Down Deriving Mechanism in Haskell (draft)}

\author{Song Zhang}
\email{Haskell.zhang.song@hotmail.com}

% \author{Anonymous}
% \email{anonymous@@mail.com}

%\author{Zirun Zhu}
%\email{zirunzhu@@gmail.com}

%% \authornote{Both authors contributed equally to this research.}

% \renewcommand{\shortauthors}{Song}

\begin{abstract}
	In Haskell, class instance deriving and related mechanisms are used pervasively. Their influence extends beyond Haskell: \textit{Rust} adopts similar ideas in its \textit{trait} system, and Java libraries such as \textit{Lombok} use annotations to generate methods such as \textit{equals} and \textit{toString}. This paper proposes an extension to Haskell's existing deriving machinery so that it can operate over composite data types and multi-level type class hierarchies. With this approach, programmers no longer need to write deriving clauses for every data declaration or explicitly enumerate all superclasses in a type class hierarchy.
\end{abstract}



\maketitle

\section{Introduction}
\subsection{Background and Related Work}
Haskell provides several mechanisms for generating class instances automatically. A tool called \textit{DrIFT}\cite{DrIFT} was built as a preprocessor that parses Haskell source code and generates instance declarations. The main Haskell compiler, \textit{GHC}, provides similar functionality internally for common classes such as \textit{Show}, \textit{Eq}, and \textit{Ord}. Later, GHC introduced language extensions that derive instances for \textit{Typeable}, \textit{Data}, \textit{Lift}, and higher-order classes such as \textit{Functor} and \textit{Foldable}.

Another way to declare a type is with the \textit{newtype} keyword. A \textit{newtype} must have exactly one constructor with exactly one field, so its runtime representation can be safely coerced\cite{DBLP:journals/corr/abs-1905-13706} to the field type. It is therefore natural to derive a class instance for the wrapper whenever the wrapped field already has that instance. This is the purpose of \textit{GeneralizedNewtypeDeriving}.

With generic programming in Haskell \cite{10.1007/11783596_14, hinze2000derivable, generic_deriving_mechanism, lmmel2003scrap, peytonjones2005scrap}, default type class methods can be defined rather than merely declared by using the \textit{DefaultSignatures} extension. Combined with \textit{DeriveAnyClass}, this allows programmers to write either an empty instance or just the class name in a deriving clause after a data type declaration. To resolve ambiguity among these deriving mechanisms, \citeauthor{deriving_strategies}\cite{deriving_strategies}\cite{tamingthezoo} introduced three deriving strategies, \textit{stock}, \textit{newtype}, and \textit{anyclass}, into GHC to denote the default mechanism, \textit{GeneralizedNewtypeDeriving}, and \textit{DeriveAnyClass}, respectively.

Meta-programming with Template Haskell \cite{sheard2002template} also facilitates class instance generation. Libraries such as \textit{aeson} provide both generic-programming and meta-programming interfaces for automatic instance generation.

\citeauthor{mitchell:derive_04_sep_2009}\cite{mitchell:derive_04_sep_2009} introduced a DSL in the \textit{derive}\footnote{This package is obsolete for GHC versions later than 8.3.} package for describing boilerplate instance code and using Template Haskell to generate type class instances such as \ensuremath{\Conid{Arbitrary}} in the \textit{QuickCheck} package or \ensuremath{\Conid{Binary}} in the \textit{binary} package.

\citeauthor{AndresLoh:2018:deriving_via}\cite{AndresLoh:2018:deriving_via} introduced a mechanism called \textit{deriving via}, which uses an intermediate type to derive class instances. Meanwhile, the \textit{deriving-compat} library uses Template Haskell to provide similar support for older GHC versions.

A substantial number of GHC language extensions exist specifically to support deriving. In GHC 9.10, 12 of the 130 language extensions are related to class instance deriving. This paper extends that line of work.

\subsection{Contribution}
Existing deriving mechanisms in GHC focus on generating one class instance for one type at a time. The primary contribution of this work is a method that enables these mechanisms to generate multiple related class instances automatically. This reduces boilerplate for Haskell programmers in two common situations:

\begin{enumerate}
\item Many data types are composite and nested together, each of which needs a deriving clause or instance declaration.
\item A data type needs to be derived as an instance of a class with many superclasses, so programmers need to enumerate all the superclasses in a deriving clause.
\end{enumerate}

The first case was proposed in GHC issue \#10607\footnote{See \url{https://gitlab.haskell.org/ghc/ghc/-/issues/10607}}. It is common among programmers who use Haskell to implement compilers and manipulate abstract syntax trees. If one wants to derive a class such as \ensuremath{\Conid{Show}}, each data type in the tree may need its own deriving clause. With the mechanism described in this paper, a single deriving declaration is sufficient.

This case requires an explicit class context, whereas deriving clauses attached to \ensuremath{\mathbf{data}} or \ensuremath{\mathbf{newtype}} declarations do not. Generating such a context for a standalone deriving or instance declaration is a secondary contribution of this paper, and it is discussed in Section \ref{context_generation}. In some cases, the generation process must stop at selected types; we discuss this in Section \ref{generation_break}. Another important feature of the context-generation algorithm is support for \textit{type families}. This support is necessary for techniques such as ``trees that grow''\cite{DBLP:journals/corr/NajdJ16}, which extend abstract syntax trees.

The second case was proposed in GHC issue \#13368\footnote{See \url{https://gitlab.haskell.org/ghc/ghc/-/issues/13368}}. Compared with the first case, it is less common, but it can still be tedious. For example, the numeric class hierarchy in GHC is carefully designed, and \ensuremath{\Conid{RealFloat}} has seven superclasses. When deriving an instance of \ensuremath{\Conid{RealFloat}}, all of those superclasses must ordinarily appear in the deriving clause. With the mechanism proposed here, a single class name is enough. Moreover, higher-order type classes of kind $(\star \rightarrow \star) \rightarrow Constraint$ are also supported, allowing classes such as \ensuremath{\Conid{MonadIO}} to be derived.

These two cases are discussed separately in Sections \ref{deriving_composite} and \ref{derive_multiple_class_instances}. For the first case, we present examples and a sketch of a general algorithm that can be used by all APIs. Section \ref{context_generation} then discusses several approaches to context generation. Programmers may also need to specify stopping conditions for instance generation and to debug the generation process; these topics are discussed at the end of Section \ref{deriving_composite}. For the second case, we again begin with examples and then present its generation algorithm, which is simpler than that of the first case.

\section{Deriving Instances for Composite Data Types} \label{deriving_composite}

Data type declarations in Haskell can be composite and deeply nested. For example, when programmers use Haskell to define an abstract syntax tree for a language, it is common to have 10 to 30 data types. The latest GHC AST contains more than 100 data types \cite{najd2017trees}. Writing all deriving clauses for a given class by hand is therefore a clear source of boilerplate. We can provide three kinds of APIs:
\begin{enumerate}
\item Standalone deriving, which generates several standalone deriving declarations.
\item Empty instances, which generate empty instance declarations for a class with default implementations based on generic programming techniques.
\item Meta-programming deriving, which applies Template Haskell functions for instance generation through APIs such as \textit{deriveJSON} in the \textit{aeson} package.
\end{enumerate}

\subsection{APIs and Examples} \label{api_and_examples}
We traverse composite data types and generate standalone deriving or instance declarations using Template Haskell. The API signatures can be defined as follows:

\begin{hscode}\SaveRestoreHook
\column{B}{@{}>{\hspre}l<{\hspost}@{}}%
\column{E}{@{}>{\hspre}l<{\hspost}@{}}%
\>[B]{}\Varid{deriving\char95 }\mathbin{::}\Conid{ClassName}\to \Conid{TypeName}\to \Conid{Q}\;[\mskip1.5mu \Conid{Dec}\mskip1.5mu]{}\<[E]%
\\
\>[B]{}\Varid{instance\char95 }\mathbin{::}\Conid{ClassName}\to \Conid{TypeName}\to \Conid{Q}\;[\mskip1.5mu \Conid{Dec}\mskip1.5mu]{}\<[E]%
\\
\>[B]{}\Varid{deriving\char95 th}\mathbin{::}(\Conid{Name},\Conid{Name}\to \Conid{Q}\;[\mskip1.5mu \Conid{Dec}\mskip1.5mu])\to \Conid{Name}\to \Conid{Q}\;[\mskip1.5mu \Conid{Dec}\mskip1.5mu]{}\<[E]%
\ColumnHook
\end{hscode}\resethooks

For example, suppose we want to derive \ensuremath{\Conid{Show}} instances for type \ensuremath{\Conid{P}} in the following code:
\begin{hscode}\SaveRestoreHook
\column{B}{@{}>{\hspre}l<{\hspost}@{}}%
\column{E}{@{}>{\hspre}l<{\hspost}@{}}%
\>[B]{}\mathbf{data}\;\Conid{P}\;\Varid{a}\;\Varid{b}\mathrel{=}\Conid{P}\;(\Conid{E}\;\Varid{a}\;(\Conid{Px}\;\Varid{b})){}\<[E]%
\\
\>[B]{}\mathbf{data}\;\Conid{E}\;\Varid{x}\;\Varid{y}\mathrel{=}\Conid{L}\;\Varid{x}\mid \Conid{R}\;\Varid{y}{}\<[E]%
\\
\>[B]{}\mathbf{data}\;\Conid{Px}\;\Varid{z}\mathrel{=}\Conid{Px}{}\<[E]%
\ColumnHook
\end{hscode}\resethooks

As shown in Figure \ref{show_generic_instance}, the \ensuremath{\Varid{deriving\char95 }} function should generate three standalone deriving declarations with the appropriate type context for each type. Here, the role\cite{DBLP:journals/corr/abs-1905-13706} of type parameter \ensuremath{\Varid{b}} in type \ensuremath{\Conid{P}} is \textit{phantom}, so it should not appear in the class context of the standalone deriving declaration for \ensuremath{\Conid{P}}. By contrast, \ensuremath{\Conid{Generic}} requires no class context at all, because it depends only on the algebraic structure of a data type.

\begin{figure}
\begin{minipage}{0.55\linewidth}
\begin{hscode}\SaveRestoreHook
\column{B}{@{}>{\hspre}l<{\hspost}@{}}%
\column{E}{@{}>{\hspre}l<{\hspost}@{}}%
\>[B]{}\Varid{deriving\char95 }\;\,$\textquotesingle\textquotesingle$\!\!\;\Conid{Show}\;\,$\textquotesingle\textquotesingle$\!\!\;\Conid{P}{}\<[E]%
\\
\>[B]{}\mathbin{======>}{}\<[E]%
\\
\>[B]{}\mathbf{deriving}\;\mathbf{instance}\;(\Conid{Show}\;\Varid{x},\Conid{Show}\;\Varid{y})\Rightarrow \Conid{Show}\;(\Conid{E}\;\Varid{x}\;\Varid{y}){}\<[E]%
\\
\>[B]{}\mathbf{deriving}\;\mathbf{instance}\;\Conid{Show}\;(\Conid{Px}\;\Varid{z}){}\<[E]%
\\
\>[B]{}\mathbf{deriving}\;\mathbf{instance}\;\Conid{Show}\;\Varid{a}\Rightarrow \Conid{Show}\;(\Conid{P}\;\Varid{a}\;\Varid{b}){}\<[E]%
\ColumnHook
\end{hscode}\resethooks
\end{minipage}\hfill
\begin{minipage}{0.4\linewidth}
\begin{hscode}\SaveRestoreHook
\column{B}{@{}>{\hspre}l<{\hspost}@{}}%
\column{E}{@{}>{\hspre}l<{\hspost}@{}}%
\>[B]{}\Varid{deriving\char95 }\;\,$\textquotesingle\textquotesingle$\!\!\;\Conid{Generic}\;\,$\textquotesingle\textquotesingle$\!\!\;\Conid{P}{}\<[E]%
\\
\>[B]{}\mathbin{======>}{}\<[E]%
\\
\>[B]{}\mathbf{deriving}\;\mathbf{instance}\;\Conid{Generic}\;(\Conid{E}\;\Varid{x}\;\Varid{y}){}\<[E]%
\\
\>[B]{}\mathbf{deriving}\;\mathbf{instance}\;\Conid{Generic}\;(\Conid{Px}\;\Varid{z}){}\<[E]%
\\
\>[B]{}\mathbf{deriving}\;\mathbf{instance}\;\Conid{Generic}\;(\Conid{P}\;\Varid{a}\;\Varid{b}){}\<[E]%
\ColumnHook
\end{hscode}\resethooks
\end{minipage}
\caption{Deriving \ensuremath{\Conid{Show}} and \ensuremath{\Conid{Generic}}}
\label{show_generic_instance}
\end{figure}

With all three types being instances of \ensuremath{\Conid{Generic}}, we can use \ensuremath{\Varid{instance\char95 }} to generate empty \ensuremath{\Conid{Binary}} instances in order to serialize them.

\begin{hscode}\SaveRestoreHook
\column{B}{@{}>{\hspre}l<{\hspost}@{}}%
\column{E}{@{}>{\hspre}l<{\hspost}@{}}%
\>[B]{}\Varid{instance\char95 }\;\,$\textquotesingle\textquotesingle$\!\!\;\Conid{Binary}\;\,$\textquotesingle\textquotesingle$\!\!\;\Conid{P}{}\<[E]%
\\
\>[B]{}\mathbin{======>}{}\<[E]%
\\
\>[B]{}\mathbf{instance}\;(\Conid{Binary}\;\Varid{x},\Conid{Binary}\;\Varid{y})\Rightarrow \Conid{Binary}\;(\Conid{E}\;\Varid{x}\;\Varid{y}){}\<[E]%
\\
\>[B]{}\mathbf{instance}\;\Conid{Binary}\;(\Conid{Px}\;\Varid{z}){}\<[E]%
\\
\>[B]{}\mathbf{instance}\;\Conid{Binary}\;\Varid{a}\Rightarrow \Conid{Binary}\;(\Conid{P}\;\Varid{a}\;\Varid{b}){}\<[E]%
\ColumnHook
\end{hscode}\resethooks

Of course, \ensuremath{\Varid{deriving\char95 }} can also be used here with the \textit{DeriveAnyClass} extension. Therefore, an API that accepts deriving strategies\cite{deriving_strategies} should be provided alongside the regular ones.

\begin{hscode}\SaveRestoreHook
\column{B}{@{}>{\hspre}l<{\hspost}@{}}%
\column{E}{@{}>{\hspre}l<{\hspost}@{}}%
\>[B]{}\Varid{strategy\char95 deriving}\mathbin{::}\Conid{DerivStrategy}\to \Conid{ClassName}\to \Conid{TypeName}\to \Conid{Q}\;[\mskip1.5mu \Conid{Dec}\mskip1.5mu]{}\<[E]%
\ColumnHook
\end{hscode}\resethooks

Then the empty instance for \ensuremath{\Conid{Binary}} can be equivalently rewritten as:

\begin{hscode}\SaveRestoreHook
\column{B}{@{}>{\hspre}l<{\hspost}@{}}%
\column{E}{@{}>{\hspre}l<{\hspost}@{}}%
\>[B]{}\Varid{strategy\char95 deriving}\;\Varid{anyclass}\;\,$\textquotesingle\textquotesingle$\!\!\;\Conid{Binary}\;\,$\textquotesingle\textquotesingle$\!\!\;\Conid{P}{}\<[E]%
\\
\>[B]{}\mathbin{======>}{}\<[E]%
\\
\>[B]{}\mathbf{deriving}\;\Varid{anyclass}\;\mathbf{instance}\;(\Conid{Binary}\;\Varid{x},\Conid{Binary}\;\Varid{y})\Rightarrow \Conid{Binary}\;(\Conid{E}\;\Varid{x}\;\Varid{y}){}\<[E]%
\\
\>[B]{}\mathbf{deriving}\;\Varid{anyclass}\;\mathbf{instance}\;\Conid{Binary}\;(\Conid{Px}\;\Varid{z}){}\<[E]%
\\
\>[B]{}\mathbf{deriving}\;\Varid{anyclass}\;\mathbf{instance}\;\Conid{Binary}\;\Varid{a}\Rightarrow \Conid{Binary}\;(\Conid{P}\;\Varid{a}\;\Varid{b}){}\<[E]%
\ColumnHook
\end{hscode}\resethooks

To generate instances with \ensuremath{\Varid{deriving\char95 th}}, a function of type \ensuremath{\Conid{Name}\to \Conid{Q}\;[\mskip1.5mu \Conid{Dec}\mskip1.5mu]} must be passed to it. Figure \ref{Arity} defines a class called \ensuremath{\Conid{Arity}}, which returns the number of type parameters of a data type. When \ensuremath{\Varid{deriving\char95 th}} is called on \ensuremath{\Conid{P}}, it expands into three instances for the types reachable from \ensuremath{\Conid{P}}. Similarly, this class does not require a class context, just like \ensuremath{\Conid{Generic}}.

\begin{figure}
\begin{minipage}[t]{0.45\linewidth}
\begin{hscode}\SaveRestoreHook
\column{B}{@{}>{\hspre}l<{\hspost}@{}}%
\column{3}{@{}>{\hspre}l<{\hspost}@{}}%
\column{5}{@{}>{\hspre}l<{\hspost}@{}}%
\column{7}{@{}>{\hspre}l<{\hspost}@{}}%
\column{9}{@{}>{\hspre}l<{\hspost}@{}}%
\column{E}{@{}>{\hspre}l<{\hspost}@{}}%
\>[B]{}\mathbf{class}\;\Conid{Arity}\;(\Varid{cla}\mathbin{::}\Varid{k})\;\mathbf{where}{}\<[E]%
\\
\>[B]{}\hsindent{5}{}\<[5]%
\>[5]{}\Varid{arity}\mathbin{::}\Conid{Proxy}\;\Varid{cla}\to \Conid{Integer}{}\<[E]%
\\[\blanklineskip]%
\>[B]{}\Varid{getArity}\mathbin{::}\Conid{Name}\to \Conid{Q}\;\Conid{Int}{}\<[E]%
\\
\>[B]{}\Varid{getArity}\;\Varid{name}\mathrel{=}\mathbf{do}{}\<[E]%
\\
\>[B]{}\hsindent{3}{}\<[3]%
\>[3]{}\Varid{info}\leftarrow \Varid{reify}\;\Varid{name}{}\<[E]%
\\
\>[B]{}\hsindent{3}{}\<[3]%
\>[3]{}\mathbf{case}\;\Varid{info}\;\mathbf{of}{}\<[E]%
\\
\>[3]{}\hsindent{2}{}\<[5]%
\>[5]{}\Conid{TyConI}\;\Varid{dec}\to \mathbf{case}\;\Varid{dec}\;\mathbf{of}{}\<[E]%
\\
\>[5]{}\hsindent{2}{}\<[7]%
\>[7]{}\Conid{DataD}\;\anonymous \;\anonymous \;\Varid{tvbs}\;\anonymous \;\anonymous \;\anonymous \to {}\<[E]%
\\
\>[7]{}\hsindent{2}{}\<[9]%
\>[9]{}\Varid{return}\mathbin{\$}\Varid{length}\;\Varid{tvbs}{}\<[E]%
\\
\>[5]{}\hsindent{2}{}\<[7]%
\>[7]{}\Conid{NewtypeD}\;\anonymous \;\anonymous \;\Varid{tvbs}\;\anonymous \;\anonymous \;\anonymous \to {}\<[E]%
\\
\>[7]{}\hsindent{2}{}\<[9]%
\>[9]{}\Varid{return}\mathbin{\$}\Varid{length}\;\Varid{tvbs}{}\<[E]%
\\
\>[3]{}\hsindent{2}{}\<[5]%
\>[5]{}\Conid{PrimTyConI}\;\anonymous \;\Varid{n}\;\anonymous \to \Varid{return}\;\Varid{n}{}\<[E]%
\ColumnHook
\end{hscode}\resethooks
\end{minipage}\hfill
\begin{minipage}[t]{0.45\linewidth}
\begin{hscode}\SaveRestoreHook
\column{B}{@{}>{\hspre}l<{\hspost}@{}}%
\column{5}{@{}>{\hspre}l<{\hspost}@{}}%
\column{9}{@{}>{\hspre}l<{\hspost}@{}}%
\column{13}{@{}>{\hspre}l<{\hspost}@{}}%
\column{E}{@{}>{\hspre}l<{\hspost}@{}}%
\>[B]{}\Varid{mkArity}\mathbin{::}\Conid{Name}\to \Conid{Q}\;[\mskip1.5mu \Conid{Dec}\mskip1.5mu]{}\<[E]%
\\
\>[B]{}\Varid{mkArity}\;\Varid{n}\mathrel{=}\mathbf{do}{}\<[E]%
\\
\>[B]{}\hsindent{5}{}\<[5]%
\>[5]{}\Varid{at}\leftarrow \Varid{getArity}\;\Varid{n}{}\<[E]%
\\
\>[B]{}\hsindent{5}{}\<[5]%
\>[5]{}[\mskip1.5mu \Varid{d}\mid \mathbf{instance}\;\Conid{Arity}\mathbin{\$}(\Varid{conT}\;\Varid{n})\;\mathbf{where}{}\<[E]%
\\
\>[5]{}\hsindent{8}{}\<[13]%
\>[13]{}\Varid{arity}\;\anonymous \mathrel{=}\Varid{at}\mid \mskip1.5mu]{}\<[E]%
\\[\blanklineskip]%
\>[B]{}\Varid{deriving\char95 th}\;(\,$\textquotesingle\textquotesingle$\!\!\;\Conid{Arity},\Varid{mkArity})\;\,$\textquotesingle\textquotesingle$\!\!\;\Conid{P}{}\<[E]%
\\
\>[B]{}\mathbin{======>}{}\<[E]%
\\
\>[B]{}\mathbf{instance}\;\Conid{Arity}\;\Conid{E}\;\mathbf{where}{}\<[E]%
\\
\>[B]{}\hsindent{9}{}\<[9]%
\>[9]{}\Varid{arity}\;\anonymous \mathrel{=}\mathrm{2}{}\<[E]%
\\
\>[B]{}\mathbf{instance}\;\Conid{Arity}\;\Conid{Px}\;\mathbf{where}{}\<[E]%
\\
\>[B]{}\hsindent{9}{}\<[9]%
\>[9]{}\Varid{arity}\;\anonymous \mathrel{=}\mathrm{1}{}\<[E]%
\\
\>[B]{}\mathbf{instance}\;\Conid{Arity}\;\Conid{P}\;\mathbf{where}{}\<[E]%
\\
\>[B]{}\hsindent{9}{}\<[9]%
\>[9]{}\Varid{arity}\;\anonymous \mathrel{=}\mathrm{2}{}\<[E]%
\ColumnHook
\end{hscode}\resethooks
\end{minipage}
\caption{The \ensuremath{\Conid{Arity}} class, its Template Haskell instance generator, and the use of \ensuremath{\Varid{deriving\char95 th}}}
\label{Arity}
\end{figure}

\subsection{Sketched Algorithm of Instance Generation} \label{generation_alg}
All three kinds of APIs share a common idea. Given a type and a class, we traverse the data types that compose it in a top-down order and generate standalone deriving or instance declarations whenever the currently visited type is not already an instance of the class. We use \ensuremath{\Conid{StateT}\;[\mskip1.5mu \Conid{Type}\mskip1.5mu]\;\Conid{Q}\;[\mskip1.5mu \Conid{Dec}\mskip1.5mu]} to generate the declarations. The \ensuremath{[\mskip1.5mu \Conid{Type}\mskip1.5mu]} state records the types for which declarations have already been generated. The general algorithm is shown in Figure \ref{general_algorithm}.
\begin{figure}
\begin{hscode}\SaveRestoreHook
\column{B}{@{}>{\hspre}l<{\hspost}@{}}%
\column{9}{@{}>{\hspre}l<{\hspost}@{}}%
\column{17}{@{}>{\hspre}l<{\hspost}@{}}%
\column{19}{@{}>{\hspre}l<{\hspost}@{}}%
\column{25}{@{}>{\hspre}l<{\hspost}@{}}%
\column{E}{@{}>{\hspre}l<{\hspost}@{}}%
\>[B]{}\mathbf{type}\;\Conid{ClassName}\mathrel{=}\Conid{Name}{}\<[E]%
\\
\>[B]{}\mathbf{type}\;\Conid{TypeName}\mathrel{=}\Conid{Name}{}\<[E]%
\\
\>[B]{}\Varid{general\char95 derive}\mathbin{::}\Conid{ClassName}\to \Conid{TypeName}\to \Conid{StateT}\;[\mskip1.5mu \Conid{Type}\mskip1.5mu]\;\Conid{Q}\;[\mskip1.5mu \Conid{Dec}\mskip1.5mu]{}\<[E]%
\\
\>[B]{}\Varid{general\char95 derive}\;\Varid{cn}\;\Varid{tn}\mathrel{=}\mathbf{do}{}\<[E]%
\\
\>[B]{}\hsindent{9}{}\<[9]%
\>[9]{}\Varid{ts}\leftarrow \Varid{get}{}\<[E]%
\\
\>[B]{}\hsindent{9}{}\<[9]%
\>[9]{}\mathbf{let}\;\Varid{t}\mathrel{=}\Varid{constructType}\;\Varid{tn}{}\<[E]%
\\
\>[B]{}\hsindent{9}{}\<[9]%
\>[9]{}\mathbf{if}\;\Varid{isInstance}\;\Varid{cn}\;\Varid{t}\mathrel{\vee}\Varid{elem}\;\Varid{t}\;\Varid{ts}\mathrel{\vee}\Varid{other\char95 cond}{}\<[E]%
\\
\>[9]{}\hsindent{8}{}\<[17]%
\>[17]{}\mathbf{then}\;\Varid{return}\;[\mskip1.5mu \mskip1.5mu]{}\<[E]%
\\
\>[9]{}\hsindent{8}{}\<[17]%
\>[17]{}\mathbf{else}\;\mathbf{do}{}\<[E]%
\\
\>[17]{}\hsindent{8}{}\<[25]%
\>[25]{}\Varid{cxt}\leftarrow \Varid{genClassContext}\;\Varid{cn}\;\Varid{tn}{}\<[E]%
\\
\>[17]{}\hsindent{8}{}\<[25]%
\>[25]{}\mathbf{let}\;\Varid{dec}\mathrel{=}\Varid{genDecl}\;\Varid{cn}\;\Varid{cxt}\;\Varid{t}{}\<[E]%
\\
\>[17]{}\hsindent{8}{}\<[25]%
\>[25]{}\Varid{modify}\;(\Varid{t}\mathbin{:}){}\<[E]%
\\
\>[17]{}\hsindent{8}{}\<[25]%
\>[25]{}\Varid{names}\leftarrow \Varid{lift}\mathbin{\$}\Varid{getCompositeNames}\;\Varid{tn}{}\<[E]%
\\
\>[17]{}\hsindent{8}{}\<[25]%
\>[25]{}\Varid{ns}\leftarrow \Varid{filterM}\;\Varid{isDataType}\;\Varid{names}{}\<[E]%
\\
\>[17]{}\hsindent{8}{}\<[25]%
\>[25]{}\Varid{decs}\leftarrow \Varid{mapM}\;(\Varid{general\char95 derive}\;\Varid{cn})\;\Varid{ns}{}\<[E]%
\\
\>[17]{}\hsindent{8}{}\<[25]%
\>[25]{}\Varid{return}\mathbin{\$}\Varid{dec}\mathbin{:}\Varid{concat}\;\Varid{decs}{}\<[E]%
\\[\blanklineskip]%
\>[B]{}\Varid{constructType}{}\<[19]%
\>[19]{}\mathbin{::}\Conid{Name}\to \Conid{Type}{}\<[E]%
\\
\>[B]{}\Varid{getCompositeNames}\mathbin{::}\Conid{Name}\to \Conid{Q}\;[\mskip1.5mu \Conid{Name}\mskip1.5mu]{}\<[E]%
\\
\>[B]{}\Varid{isDataType}{}\<[19]%
\>[19]{}\mathbin{::}\Conid{Name}\to \Conid{Q}\;\Conid{Bool}{}\<[E]%
\\
\>[B]{}\Varid{genClassContext}{}\<[19]%
\>[19]{}\mathbin{::}\Conid{ClassName}\to \Conid{TypeName}\to \Conid{Q}\;\Conid{Cxt}{}\<[E]%
\\
\>[B]{}\Varid{genDecl}{}\<[19]%
\>[19]{}\mathbin{::}\Conid{ClassName}\to \Conid{Cxt}\to \Conid{Type}\to \Conid{Dec}{}\<[E]%
\ColumnHook
\end{hscode}\resethooks
\caption{General top-down deriving algorithm}
\label{general_algorithm}
\end{figure}

The algorithm first checks whether the current type is already an instance or has already been generated. If either condition holds, the generation process stops at that point. The third condition, \ensuremath{\Varid{other\char95 cond}}, is discussed in detail in Section \ref{generation_break}. Otherwise, we generate a declaration. Before doing so, we construct its class context with \ensuremath{\Varid{genClassContext}}, which is discussed in Section \ref{context_generation}. We then recursively apply the generation function to the type names that compose the current type. These names can be obtained with \ensuremath{\Varid{getCompositeNames}}. Type variables and type families must be excluded from its result, since they may appear in the context but cannot themselves serve as instance heads.

\subsection{\ensuremath{\Varid{isInstance}} Function for Polymorphic Types}
As seen in the algorithm, it relies on checking whether a type is an instance of a class. However, the \textit{template-haskell} function \ensuremath{\Varid{isInstance}} does not support polymorphic types: whether or not we provide a type context, both examples in Figure \ref{polymorphic_types} return \ensuremath{\Conid{False}}.

\begin{figure}
\begin{minipage}{0.45\linewidth}
\begin{hscode}\SaveRestoreHook
\column{B}{@{}>{\hspre}l<{\hspost}@{}}%
\column{5}{@{}>{\hspre}l<{\hspost}@{}}%
\column{E}{@{}>{\hspre}l<{\hspost}@{}}%
\>[B]{}\Varid{poly\char95 a}\mathbin{::}\Conid{Q}\;\Conid{Bool}{}\<[E]%
\\
\>[B]{}\Varid{poly\char95 a}\mathrel{=}\mathbf{do}{}\<[E]%
\\
\>[B]{}\hsindent{5}{}\<[5]%
\>[5]{}\Varid{poly\char95 a\char95 t}\leftarrow [\mskip1.5mu \Varid{t}\mid \Varid{forall}\;\Varid{a}.\Conid{Eq}\;\Varid{a}\Rightarrow [\mskip1.5mu \Varid{a}\mskip1.5mu]\mid \mskip1.5mu]{}\<[E]%
\\
\>[B]{}\hsindent{5}{}\<[5]%
\>[5]{}\Varid{isInstance}\;\,$\textquotesingle\textquotesingle$\!\!\;\Conid{Eq}\;[\mskip1.5mu \Varid{poly\char95 a\char95 t}\mskip1.5mu]{}\<[E]%
\ColumnHook
\end{hscode}\resethooks
\end{minipage}\hfill
\begin{minipage}{0.45\linewidth}
\begin{hscode}\SaveRestoreHook
\column{B}{@{}>{\hspre}l<{\hspost}@{}}%
\column{5}{@{}>{\hspre}l<{\hspost}@{}}%
\column{E}{@{}>{\hspre}l<{\hspost}@{}}%
\>[B]{}\Varid{poly\char95 a'}\mathbin{::}\Conid{Q}\;\Conid{Bool}{}\<[E]%
\\
\>[B]{}\Varid{poly\char95 a'}\mathrel{=}\mathbf{do}{}\<[E]%
\\
\>[B]{}\hsindent{5}{}\<[5]%
\>[5]{}\Varid{poly\char95 a\char95 t}\leftarrow [\mskip1.5mu \Varid{t}\mid \Varid{forall}\;\Varid{a}.[\mskip1.5mu \Varid{a}\mskip1.5mu]\mid \mskip1.5mu]{}\<[E]%
\\
\>[B]{}\hsindent{5}{}\<[5]%
\>[5]{}\Varid{isInstance}\;\,$\textquotesingle\textquotesingle$\!\!\;\Conid{Eq}\;[\mskip1.5mu \Varid{poly\char95 a\char95 t}\mskip1.5mu]{}\<[E]%
\ColumnHook
\end{hscode}\resethooks
\end{minipage}

\caption{\ensuremath{\Varid{isInstance}} for polymorphic types}
\label{polymorphic_types}
\end{figure}

To check polymorphic types correctly, Scott proposed replacing all type variables with \ensuremath{\Conid{Any}} from \ensuremath{\Conid{\Conid{GHC}.Exts}}\footnote{See \url{https://gitlab.haskell.org/ghc/ghc/-/issues/10607\#note_132897}}. This transformation can be implemented easily with \textit{syb} generic programming\cite{lmmel2003scrap}. In practice, we define a transformation that replaces every type variable with \ensuremath{\Conid{Any}} and apply it throughout the type. In other words, instead of checking whether \ensuremath{[\mskip1.5mu \Varid{a}\mskip1.5mu]} or \ensuremath{\Varid{forall}\;\Varid{a}.\Conid{Eq}\;\Varid{a}\Rightarrow [\mskip1.5mu \Varid{a}\mskip1.5mu]} is an instance of \ensuremath{\Conid{Eq}}, we check whether \ensuremath{[\mskip1.5mu \Conid{Any}\mskip1.5mu]} is. Throughout the rest of this paper, the \ensuremath{\Varid{isInstance}} function should therefore be understood as a reimplementation that applies this \ensuremath{\Conid{Any}} transformation.

\section{Context Generation} \label{context_generation}
There are many ways to provide the class context for an instance in modern GHC. In this section, we discuss two simple context generation methods and then propose an algorithm for inferring the class context of an instance. We also briefly note other possible approaches at the end of the section. Before discussing context generation itself, however, we first extend our deriving APIs so that they can work with arbitrary context generation functions.

\subsection{Extended APIs}
We extend the APIs by parameterizing the \ensuremath{\Varid{genClassContext}} function in the generation algorithm. The type of the new \ensuremath{\Varid{general\char95 derive}} can be defined as:
\begin{hscode}\SaveRestoreHook
\column{B}{@{}>{\hspre}l<{\hspost}@{}}%
\column{E}{@{}>{\hspre}l<{\hspost}@{}}%
\>[B]{}\mathbf{type}\;\Conid{ContextGenerator}\mathrel{=}\Conid{ClassName}\to \Conid{TypeName}\to \Conid{Q}\;\Conid{Cxt}{}\<[E]%
\\
\>[B]{}\Varid{general\char95 derive}\mathbin{::}\Conid{ClassName}\to \Conid{TypeName}\to \Conid{ContextGenerator}\to \Conid{StateT}\;[\mskip1.5mu \Conid{Type}\mskip1.5mu]\;\Conid{Q}\;[\mskip1.5mu \Conid{Dec}\mskip1.5mu]{}\<[E]%
\ColumnHook
\end{hscode}\resethooks

Both \ensuremath{\Varid{deriving\char95 }} and \ensuremath{\Varid{instance\char95 }} must also be extended with a parameter of type \ensuremath{\Conid{ContextGenerator}} so that they can accommodate different context inference algorithms. By contrast, \ensuremath{\Varid{deriving\char95 th}} does not need this parameter, because the generation of instance declarations is handled entirely by the meta function of type \ensuremath{\Conid{Name}\to \Conid{Q}\;[\mskip1.5mu \Conid{Dec}\mskip1.5mu]}.
\begin{hscode}\SaveRestoreHook
\column{B}{@{}>{\hspre}l<{\hspost}@{}}%
\column{E}{@{}>{\hspre}l<{\hspost}@{}}%
\>[B]{}\Varid{deriving\char95 with}\mathbin{::}\Conid{ClassName}\to \Conid{TypeName}\to \Conid{ContextGenerator}\to \Conid{Q}\;[\mskip1.5mu \Conid{Dec}\mskip1.5mu]{}\<[E]%
\\
\>[B]{}\Varid{instance\char95 with}\mathbin{::}\Conid{ClassName}\to \Conid{TypeName}\to \Conid{ContextGenerator}\to \Conid{Q}\;[\mskip1.5mu \Conid{Dec}\mskip1.5mu]{}\<[E]%
\ColumnHook
\end{hscode}\resethooks
 
\subsection{Hole as a Context}
The simplest approach is to use a wildcard with the \textit{PartialTypeSignatures} extension, as proposed by Scott\footnote{See \url{https://gitlab.haskell.org/ghc/ghc/-/issues/13324}}. In many cases, GHC can infer the context of a standalone deriving or instance declaration by itself. The limitation of this method is that it may not handle cases in which type families appear in the data type.
The implementation of this context generation function is very simple:

\begin{hscode}\SaveRestoreHook
\column{B}{@{}>{\hspre}l<{\hspost}@{}}%
\column{E}{@{}>{\hspre}l<{\hspost}@{}}%
\>[B]{}\Varid{genHoleContext}\mathbin{::}\Conid{ClassName}\to \Conid{TypeName}\to \Conid{Q}\;\Conid{Cxt}{}\<[E]%
\\
\>[B]{}\Varid{genHoleContext}\;\anonymous \;\anonymous \mathrel{=}\Varid{return}\;[\mskip1.5mu \Conid{WildCardT}\mskip1.5mu]{}\<[E]%
\ColumnHook
\end{hscode}\resethooks

We can simply pass it to \ensuremath{\Varid{deriving\char95 with}} to generate a wildcard hole as the class context:

\begin{hscode}\SaveRestoreHook
\column{B}{@{}>{\hspre}l<{\hspost}@{}}%
\column{E}{@{}>{\hspre}l<{\hspost}@{}}%
\>[B]{}\Varid{deriving\char95 with}\;\,$\textquotesingle\textquotesingle$\!\!\;\Conid{Eq}\;\,$\textquotesingle\textquotesingle$\!\!\;\Conid{P}\;\Varid{genHoleContext}{}\<[E]%
\\
\>[B]{}\mathbin{======>}{}\<[E]%
\\
\>[B]{}\mathbf{deriving}\;\mathbf{instance}\;\anonymous \Rightarrow \Conid{Eq}\;(\Conid{E}\;\Varid{x}\;\Varid{y}){}\<[E]%
\\
\>[B]{}\mathbf{deriving}\;\mathbf{instance}\;\anonymous \Rightarrow \Conid{Eq}\;(\Conid{Px}\;\Varid{z}){}\<[E]%
\\
\>[B]{}\mathbf{deriving}\;\mathbf{instance}\;\anonymous \Rightarrow \Conid{Eq}\;(\Conid{P}\;\Varid{a}\;\Varid{b}){}\<[E]%
\ColumnHook
\end{hscode}\resethooks

\subsection{All Fields as a Context}
Using all fields as a class context means collecting all field types in a data type and assembling them into the class context. This approach can even support recursive data types and data types that use type families.

\begin{hscode}\SaveRestoreHook
\column{B}{@{}>{\hspre}l<{\hspost}@{}}%
\column{E}{@{}>{\hspre}l<{\hspost}@{}}%
\>[B]{}\mathbf{data}\;\Conid{List}\;\Varid{a}\mathrel{=}\Conid{Nil}\mid \Conid{Cons}\;\Varid{a}\;(\Conid{List}\;\Varid{a}){}\<[E]%
\\
\>[B]{}\mathbf{deriving}\;\mathbf{instance}\;(\Conid{Show}\;\Varid{a},\Conid{Show}\;(\Conid{List}\;\Varid{a}))\Rightarrow \Conid{Show}\;(\Conid{List}\;\Varid{a}){}\<[E]%
\ColumnHook
\end{hscode}\resethooks

It also works for types with type families, since modern GHC is often clever enough to resolve the resulting context constraints.

\begin{hscode}\SaveRestoreHook
\column{B}{@{}>{\hspre}l<{\hspost}@{}}%
\column{E}{@{}>{\hspre}l<{\hspost}@{}}%
\>[B]{}\mathbf{type}\;\Varid{family}\;\Conid{F}\;\Varid{a}{}\<[E]%
\\
\>[B]{}\mathbf{type}\;\mathbf{instance}\;\Conid{F}\;\Conid{Int}\mathrel{=}\Conid{Char}{}\<[E]%
\\
\>[B]{}\mathbf{data}\;\Conid{MkT}\;\Varid{a}\;\Varid{b}\mathrel{=}\Conid{Mk1}\;(\Conid{F}\;\Varid{a})\;\Conid{Int}\;(\Conid{Either}\;\Conid{Int}\;\Varid{b}){}\<[E]%
\\[\blanklineskip]%
\>[B]{}\mathbf{deriving}\;\mathbf{instance}\;(\Conid{Show}\;(\Conid{F}\;\Varid{a}),\Conid{Show}\;\Conid{Int},\Conid{Show}\;(\Conid{Either}\;\Conid{Int}\;\Varid{b}))\Rightarrow \Conid{Show}\;(\Conid{MkT}\;\Varid{a}\;\Varid{b}){}\<[E]%
\ColumnHook
\end{hscode}\resethooks
This method works in most cases with language extensions such as \textit{FlexibleContexts} and \textit{UndecidableInstances}, but it is not a particularly clean solution. Even if we avoid including nullary data types such as \ensuremath{\Conid{Int}} or \ensuremath{\Conid{Bool}}, the resulting context still contains many unnecessary constraints. The process of collecting field types with Template Haskell is straightforward, so we omit it here.

\subsection{Inferring Context}
Figure \ref{syntax} presents a simplified Haskell syntax for data type declarations. We do not consider \textit{newtype}, since in this mechanism it can be treated as a special case of a declaration written with the \textit{data} keyword. \textit{GADT}s and existential types are also excluded. We classify monotypes into five categories: nullary types such as \textit{Int}, type variables \textit{v}, type family applications ${F\ [t]}$, data type applications $T\ [t]$, and type-variable applications $v\ [t]$. When inferring a class context, nullary types such as \textit{Int} can be ignored. Type variables, type family applications, and type-variable applications should be placed directly into the context. For a data type application, we recursively infer its type context and substitute its type parameters with the applied arguments. We then replace the original data type application with the result of that substitution. Finally, we repeat the process until no data type applications remain in the context.

\begin{figure}
\begin{minipage}{0.59\linewidth}
\begin{alignat*}{3}
	& type\ variables\   &   v & ::= a, b, c \cdots \\
	& data\ constructors\ & Con & ::= MkA, MkB \cdots \\
	& type\ families\      &   F & ::= F1, F2 \cdots \\
	& monotypes\       &     t & ::= Int\ \vert\ v\ \vert\ F\ \big[t\big]\ \vert\ T\ \big[t\big]\ \vert\ v\ \big[t\big]\\
	& data\ declarations\   & A,B,T & ::= data\ T\ \big[v\big] = \big[Con\ \big[t\big]\big]
\end{alignat*}
\end{minipage}\hfill
\begin{minipage}{0.39\linewidth}
\begin{alignat*}{2}
I & ::= \epsilon\ \vert\ T,\ I\\
P & ::= \epsilon\ \vert\ \big(T,\ \big[p\big]\big),\ P\\
F & ::= \epsilon\ \vert\ \big(T,\ \big[t\big]\big),\ F\\
S & ::= \epsilon\ \vert\ \big(T,\ Map\ t\ \big[\big(v,t\big)\big]\big),\ S\\
R & ::= \epsilon\ \vert\ \big(T,\ \big[t\big]\big), R
\end{alignat*}
\end{minipage}
	\centering
	\theverbbox
	\caption{Syntax and States}
	\label{syntax}
\end{figure}

The algorithm uses five pieces of state:
\begin{enumerate}
\item $I$: stores the types currently being inferred. This is used to handle recursive data types, and it is represented as a list of type names.
\item $P$: maps each type name to its parameter list. This state can be initialized before inference begins.
\item $F$: maps each type name to a list of types. When no data type applications remain in the list, we move the corresponding key-value pair into $R$.
\item $S$: stores substitutions and tracks the substitutions induced by fields in a data type.
\item $R$: stores inferred type contexts so that duplicate inference can be avoided.
\end{enumerate}

This process is more imperative than functional, so we use concrete examples to illustrate it\footnote{See \url{https://hackage.haskell.org/package/derive-topdown} for an implementation}.
\subsubsection{Non-Recursive Examples}
We take type \ensuremath{\Conid{P}} from Section \ref{api_and_examples} as an example.

First, state $P$ can be initialized as a map from type names to their type parameters.

We then put the type name \ensuremath{\Conid{P}} into $I$ and its fields into $F$. Since \ensuremath{\Conid{E}\;\Varid{a}\;(\Conid{Px}\;\Varid{b})} is a data type application, we construct a substitution map from \ensuremath{\Conid{E}}'s parameters to the arguments of that application. After that, we enumerate the data type applications inside \ensuremath{\Conid{P}} and recursively infer their class contexts. In this example, the only such type is \ensuremath{\Conid{E}}, so we put \ensuremath{\Conid{E}} into $I$ and its fields into $F$. These steps are shown in Figure \ref{infer_p_e}.

\begin{figure}
\begin{minipage}{0.48\linewidth}
\begin{hscode}\SaveRestoreHook
\column{B}{@{}>{\hspre}l<{\hspost}@{}}%
\column{E}{@{}>{\hspre}l<{\hspost}@{}}%
\>[B]{}\Conid{I}\mathrel{=}\{\mskip1.5mu \Conid{P}\mskip1.5mu\}{}\<[E]%
\\
\>[B]{}\Conid{P}\mathrel{=}\{\mskip1.5mu \Conid{P}\to [\mskip1.5mu \Varid{a},\Varid{b}\mskip1.5mu],\Conid{E}\to [\mskip1.5mu \Varid{x},\Varid{y}\mskip1.5mu],\Conid{Px}\to [\mskip1.5mu \Varid{z}\mskip1.5mu]\mskip1.5mu\}{}\<[E]%
\\
\>[B]{}\Conid{F}\mathrel{=}\{\mskip1.5mu \Conid{P}\to [\mskip1.5mu \Conid{E}\;\Varid{a}\;(\Conid{Px}\;\Varid{b})\mskip1.5mu]\mskip1.5mu\}{}\<[E]%
\\
\>[B]{}\Conid{S}\mathrel{=}\{\mskip1.5mu \Conid{P}\to \Conid{E}\;\Varid{a}\;(\Conid{Px}\;\Varid{b})\to \{\mskip1.5mu \Varid{x}\to \Varid{a},\Varid{y}\to \Conid{Px}\;\Varid{b}\mskip1.5mu\}\mskip1.5mu\}{}\<[E]%
\\
\>[B]{}\Conid{R}\mathrel{=}\{\mskip1.5mu \mskip1.5mu\}{}\<[E]%
\ColumnHook
\end{hscode}\resethooks
\end{minipage}\hfill
\begin{minipage}{0.48\linewidth}
\begin{hscode}\SaveRestoreHook
\column{B}{@{}>{\hspre}l<{\hspost}@{}}%
\column{E}{@{}>{\hspre}l<{\hspost}@{}}%
\>[B]{}\Conid{I}\mathrel{=}\{\mskip1.5mu \Conid{E},\Conid{P}\mskip1.5mu\}{}\<[E]%
\\
\>[B]{}\Conid{P}\mathrel{=}\{\mskip1.5mu \Conid{P}\to [\mskip1.5mu \Varid{a},\Varid{b}\mskip1.5mu],\Conid{E}\to [\mskip1.5mu \Varid{x},\Varid{y}\mskip1.5mu],\Conid{Px}\to [\mskip1.5mu \Varid{z}\mskip1.5mu]\mskip1.5mu\}{}\<[E]%
\\
\>[B]{}\Conid{F}\mathrel{=}\{\mskip1.5mu \Conid{P}\to [\mskip1.5mu \Conid{E}\;\Varid{a}\;(\Conid{Px}\;\Varid{b})\mskip1.5mu],\Conid{E}\to [\mskip1.5mu \Varid{x},\Varid{y}\mskip1.5mu]\mskip1.5mu\}{}\<[E]%
\\
\>[B]{}\Conid{S}\mathrel{=}\{\mskip1.5mu \Conid{P}\to \Conid{E}\;\Varid{a}\;(\Conid{Px}\;\Varid{b})\to \{\mskip1.5mu \Varid{x}\to \Varid{a},\Varid{y}\to \Conid{Px}\;\Varid{b}\mskip1.5mu\}\mskip1.5mu\}{}\<[E]%
\\
\>[B]{}\Conid{R}\mathrel{=}\{\mskip1.5mu \mskip1.5mu\}{}\<[E]%
\ColumnHook
\end{hscode}\resethooks
\end{minipage}

\caption{Inferring the class context of \ensuremath{\Conid{P}}}
\label{infer_p_e}
\end{figure}

Since the fields of \ensuremath{\Conid{E}} contain no data type applications and consist only of type variables, inference for \ensuremath{\Conid{E}} finishes here and we move it into $R$, as shown in Figure \ref{infer_e}. This condition is the base case of the algorithm.

\begin{figure}
\begin{minipage}{0.48\linewidth}
\begin{hscode}\SaveRestoreHook
\column{B}{@{}>{\hspre}l<{\hspost}@{}}%
\column{E}{@{}>{\hspre}l<{\hspost}@{}}%
\>[B]{}\Conid{I}\mathrel{=}\{\mskip1.5mu \Conid{E},\Conid{P}\mskip1.5mu\}{}\<[E]%
\\
\>[B]{}\Conid{P}\mathrel{=}\{\mskip1.5mu \Conid{P}\to [\mskip1.5mu \Varid{a},\Varid{b}\mskip1.5mu],\Conid{E}\to [\mskip1.5mu \Varid{x},\Varid{y}\mskip1.5mu],\Conid{Px}\to [\mskip1.5mu \Varid{z}\mskip1.5mu]\mskip1.5mu\}{}\<[E]%
\\
\>[B]{}\Conid{F}\mathrel{=}\{\mskip1.5mu \Conid{P}\to [\mskip1.5mu \Conid{E}\;\Varid{a}\;(\Conid{Px}\;\Varid{b})\mskip1.5mu]\mskip1.5mu\}{}\<[E]%
\\
\>[B]{}\Conid{S}\mathrel{=}\{\mskip1.5mu \Conid{P}\to \Conid{E}\;\Varid{a}\;(\Conid{Px}\;\Varid{b})\to \{\mskip1.5mu \Varid{x}\to \Varid{a},\Varid{y}\to \Conid{Px}\;\Varid{b}\mskip1.5mu\}\mskip1.5mu\}{}\<[E]%
\\
\>[B]{}\Conid{R}\mathrel{=}\{\mskip1.5mu \Conid{E}\to [\mskip1.5mu \Varid{x},\Varid{y}\mskip1.5mu]\mskip1.5mu\}{}\<[E]%
\ColumnHook
\end{hscode}\resethooks
\end{minipage}\hfill
\begin{minipage}{0.48\linewidth}
\begin{hscode}\SaveRestoreHook
\column{B}{@{}>{\hspre}l<{\hspost}@{}}%
\column{E}{@{}>{\hspre}l<{\hspost}@{}}%
\>[B]{}\Conid{I}\mathrel{=}\{\mskip1.5mu \Conid{E},\Conid{P}\mskip1.5mu\}{}\<[E]%
\\
\>[B]{}\Conid{P}\mathrel{=}\{\mskip1.5mu \Conid{P}\to [\mskip1.5mu \Varid{a},\Varid{b}\mskip1.5mu],\Conid{E}\to [\mskip1.5mu \Varid{x},\Varid{y}\mskip1.5mu],\Conid{Px}\to [\mskip1.5mu \Varid{z}\mskip1.5mu]\mskip1.5mu\}{}\<[E]%
\\
\>[B]{}\Conid{F}\mathrel{=}\{\mskip1.5mu \Conid{P}\to [\mskip1.5mu \Varid{a},\Conid{Px}\;\Varid{b}\mskip1.5mu]\mskip1.5mu\}{}\<[E]%
\\
\>[B]{}\Conid{S}\mathrel{=}\{\mskip1.5mu \Conid{P}\to \{\mskip1.5mu \Conid{E}\;\Varid{a}\;(\Conid{Px}\;\Varid{b})\to \{\mskip1.5mu \Varid{x}\to \Varid{a},\Varid{y}\to \Conid{Px}\;\Varid{b}\mskip1.5mu\}\mskip1.5mu\}\mskip1.5mu\}{}\<[E]%
\\
\>[B]{}\Conid{R}\mathrel{=}\{\mskip1.5mu \Conid{E}\to [\mskip1.5mu \Varid{x},\Varid{y}\mskip1.5mu]\mskip1.5mu\}{}\<[E]%
\ColumnHook
\end{hscode}\resethooks
\end{minipage}
\caption{Inferring the class context of \ensuremath{\Conid{E}}}
\label{infer_e}
\end{figure}

Now we apply the substitution \ensuremath{\{\mskip1.5mu \Varid{x}\to \Varid{a},\Varid{y}\to \Conid{Px}\;\Varid{b}\mskip1.5mu\}} to \ensuremath{\Conid{E}}'s class context \ensuremath{[\mskip1.5mu \Varid{x},\Varid{y}\mskip1.5mu]} and obtain \ensuremath{\{\mskip1.5mu \Varid{a},\Conid{Px}\;\Varid{b}\mskip1.5mu\}}. We then use this result to replace \ensuremath{\Conid{E}\;\Varid{a}\;(\Conid{Px}\;\Varid{b})} in the field list of \ensuremath{\Conid{P}} in $F$. The resulting state appears in Figure \ref{sub_e}.

For type \ensuremath{\Conid{Px}}, the process is the same as for \ensuremath{\Conid{E}}. We construct the map \ensuremath{\{\mskip1.5mu \Varid{z}\to \Varid{b}\mskip1.5mu\}} for the type \ensuremath{\Conid{P}} and put the result into $R$, as shown in Figure \ref{sub_e}.

\begin{figure}
\begin{hscode}\SaveRestoreHook
\column{B}{@{}>{\hspre}l<{\hspost}@{}}%
\column{E}{@{}>{\hspre}l<{\hspost}@{}}%
\>[B]{}\Conid{I}\mathrel{=}\{\mskip1.5mu \Conid{Px},\Conid{E},\Conid{P}\mskip1.5mu\}{}\<[E]%
\\
\>[B]{}\Conid{P}\mathrel{=}\{\mskip1.5mu \Conid{P}\to [\mskip1.5mu \Varid{a},\Varid{b}\mskip1.5mu],\Conid{E}\to [\mskip1.5mu \Varid{x},\Varid{y}\mskip1.5mu],\Conid{Px}\to [\mskip1.5mu \Varid{z}\mskip1.5mu]\mskip1.5mu\}{}\<[E]%
\\
\>[B]{}\Conid{F}\mathrel{=}\{\mskip1.5mu \Conid{P}\to [\mskip1.5mu \Varid{a},\Conid{Px}\;\Varid{b}\mskip1.5mu]\mskip1.5mu\}{}\<[E]%
\\
\>[B]{}\Conid{S}\mathrel{=}\{\mskip1.5mu \Conid{P}\to \{\mskip1.5mu \Conid{E}\;\Varid{a}\;(\Conid{Px}\;\Varid{b})\to \{\mskip1.5mu \Varid{x}\to \Varid{a},\Varid{y}\to \Conid{Px}\;\Varid{b}\mskip1.5mu\},\Conid{Px}\;\Varid{b}\to \{\mskip1.5mu \Varid{z}\to \Varid{b}\mskip1.5mu\}\mskip1.5mu\}\mskip1.5mu\}{}\<[E]%
\\
\>[B]{}\Conid{R}\mathrel{=}\{\mskip1.5mu \Conid{E}\to [\mskip1.5mu \Varid{x},\Varid{y}\mskip1.5mu],\Conid{Px}\to [\mskip1.5mu \mskip1.5mu]\mskip1.5mu\}{}\<[E]%
\ColumnHook
\end{hscode}\resethooks
\caption{Inferring the class context of \ensuremath{\Conid{Px}}}
\label{sub_e}
\end{figure}

Since the class context of \ensuremath{\Conid{Px}} is empty, \ensuremath{\Conid{Px}\;\Varid{b}} is replaced by the empty list, leaving only \ensuremath{\Varid{a}}, as shown in Figure \ref{finish_px}.

\begin{figure}
\begin{hscode}\SaveRestoreHook
\column{B}{@{}>{\hspre}l<{\hspost}@{}}%
\column{E}{@{}>{\hspre}l<{\hspost}@{}}%
\>[B]{}\Conid{I}\mathrel{=}\{\mskip1.5mu \Conid{Px},\Conid{E},\Conid{P}\mskip1.5mu\}{}\<[E]%
\\
\>[B]{}\Conid{P}\mathrel{=}\{\mskip1.5mu \Conid{P}\to [\mskip1.5mu \Varid{a},\Varid{b}\mskip1.5mu],\Conid{E}\to [\mskip1.5mu \Varid{x},\Varid{y}\mskip1.5mu],\Conid{Px}\to [\mskip1.5mu \Varid{z}\mskip1.5mu],\Conid{Px}\to [\mskip1.5mu \Varid{z}\mskip1.5mu]\mskip1.5mu\}{}\<[E]%
\\
\>[B]{}\Conid{F}\mathrel{=}\{\mskip1.5mu \Conid{P}\to [\mskip1.5mu \Varid{a}\mskip1.5mu]\mskip1.5mu\}{}\<[E]%
\\
\>[B]{}\Conid{S}\mathrel{=}\{\mskip1.5mu \Conid{P}\to \{\mskip1.5mu \Conid{E}\;\Varid{a}\;(\Conid{Px}\;\Varid{b})\to \{\mskip1.5mu \Varid{x}\to \Varid{a},\Varid{y}\to \Conid{Px}\;\Varid{b}\mskip1.5mu\},\Conid{Px}\;\Varid{b}\to \{\mskip1.5mu \Varid{z}\to \Varid{b}\mskip1.5mu\}\mskip1.5mu\}\mskip1.5mu\}{}\<[E]%
\\
\>[B]{}\Conid{R}\mathrel{=}\{\mskip1.5mu \Conid{E}\to [\mskip1.5mu \Varid{x},\Varid{y}\mskip1.5mu],\Conid{Px}\to [\mskip1.5mu \mskip1.5mu]\mskip1.5mu\}{}\<[E]%
\ColumnHook
\end{hscode}\resethooks
\caption{Substitute \ensuremath{[\mskip1.5mu \mskip1.5mu]} with \ensuremath{\Varid{z}\to \Varid{b}}}
\label{finish_px}
\end{figure}
Finally, once no data type applications remain in \ensuremath{\Conid{P}}'s context, we move it into $R$ and return it. The intermediate states are straightforward and are therefore omitted.

\subsubsection{Recursive Examples}
The example above is relatively simple because it only requires expansion and substitution of data type applications in $F$. To see how recursion is handled, consider the following recursive type:
\begin{hscode}\SaveRestoreHook
\column{B}{@{}>{\hspre}l<{\hspost}@{}}%
\column{E}{@{}>{\hspre}l<{\hspost}@{}}%
\>[B]{}\mathbf{data}\;\Conid{Stream}\;\Varid{k}\;\Varid{a}\;\Varid{b}\;\Varid{c}\mathrel{=}\Conid{S}\;(\Varid{k}\;\Varid{a})\mid \Conid{S2}\;\Varid{a}\;(\Conid{Stream}\;\Varid{k}\;\Varid{b}\;\Varid{c}\;\Conid{Int}){}\<[E]%
\ColumnHook
\end{hscode}\resethooks

As before, we put \ensuremath{\Conid{Stream}} into $I$ and place its fields, \ensuremath{\Varid{k}\;\Varid{a}} and \ensuremath{\Conid{Stream}\;\Varid{k}\;\Varid{b}\;\Varid{c}\;\Conid{Int}}, into $F$, as shown on the left side of Figure \ref{recursive_stream}. When we enumerate data type applications, we detect the recursive occurrence of \ensuremath{\Conid{Stream}} by checking whether it already appears in $I$. We then construct the substitution and remove \ensuremath{\Conid{Stream}\;\Varid{k}\;\Varid{b}\;\Varid{c}\;\Conid{Int}} without placing \ensuremath{\Conid{Stream}}'s fields into $F$ again. The substitution is shown in state $S$ on the right side of Figure \ref{recursive_stream}. Finally, we apply the substitution \ensuremath{\{\mskip1.5mu \Varid{k}\to \Varid{k},\Varid{a}\to \Varid{b},\Varid{b}\to \Varid{c},\Varid{c}\to \Conid{Int}\mskip1.5mu\}} to \ensuremath{[\mskip1.5mu \Varid{k},\Varid{a},\Varid{b},\Varid{c}\mskip1.5mu]} until a fixed point is reached. The result is \ensuremath{\{\mskip1.5mu \Varid{k}\;\Varid{a},\Varid{k}\;\Varid{b},\Varid{k}\;\Varid{c},\Varid{k}\;\Conid{Int},\Varid{a},\Varid{b},\Varid{c}\mskip1.5mu\}}. The key point is that we never put the fields of a data type into $F$ twice; when a recursive occurrence is encountered, we record only the substitution. The final step of moving the result into $R$ is omitted.

\begin{figure}
\begin{minipage}{0.45\linewidth}
\begin{hscode}\SaveRestoreHook
\column{B}{@{}>{\hspre}l<{\hspost}@{}}%
\column{E}{@{}>{\hspre}l<{\hspost}@{}}%
\>[B]{}\Conid{I}\mathrel{=}\{\mskip1.5mu \Conid{Stream}\mskip1.5mu\}{}\<[E]%
\\
\>[B]{}\Conid{P}\mathrel{=}\{\mskip1.5mu \Conid{Stream}\to [\mskip1.5mu \Varid{k},\Varid{a},\Varid{b},\Varid{c}\mskip1.5mu]\mskip1.5mu\}{}\<[E]%
\\
\>[B]{}\Conid{F}\mathrel{=}\{\mskip1.5mu \Conid{Stream}\to \{\mskip1.5mu \Varid{k}\;\Varid{a},\Conid{Stream}\;\Varid{k}\;\Varid{b}\;\Varid{c}\;\Conid{Int}\mskip1.5mu\}\mskip1.5mu\}{}\<[E]%
\\
\>[B]{}\Conid{S}\mathrel{=}\{\mskip1.5mu \mskip1.5mu\}{}\<[E]%
\\
\>[B]{}\Conid{R}\mathrel{=}\{\mskip1.5mu \mskip1.5mu\}{}\<[E]%
\ColumnHook
\end{hscode}\resethooks
\end{minipage}\hfill
\begin{minipage}{0.55\linewidth}
\begin{hscode}\SaveRestoreHook
\column{B}{@{}>{\hspre}l<{\hspost}@{}}%
\column{25}{@{}>{\hspre}l<{\hspost}@{}}%
\column{E}{@{}>{\hspre}l<{\hspost}@{}}%
\>[B]{}\Conid{I}\mathrel{=}\{\mskip1.5mu \Conid{Stream}\mskip1.5mu\}{}\<[E]%
\\
\>[B]{}\Conid{P}\mathrel{=}\{\mskip1.5mu \Conid{Stream}\to [\mskip1.5mu \Varid{k},\Varid{a},\Varid{b},\Varid{c}\mskip1.5mu]\mskip1.5mu\}{}\<[E]%
\\
\>[B]{}\Conid{F}\mathrel{=}\{\mskip1.5mu \Conid{Stream}\to \{\mskip1.5mu \Varid{k}\;\Varid{a}\mskip1.5mu\}\mskip1.5mu\}{}\<[E]%
\\
\>[B]{}\Conid{S}\mathrel{=}\{\mskip1.5mu \Conid{Stream}\to \{\mskip1.5mu (\Conid{Stream}\;\Varid{k}\;\Varid{b}\;\Varid{c}\;\Conid{Int})\to \{\mskip1.5mu \Varid{k}\to \Varid{k},{}\<[E]%
\\
\>[B]{}\hsindent{25}{}\<[25]%
\>[25]{}\Varid{a}\to \Varid{b},\Varid{b}\to \Varid{c},\Varid{c}\to \Conid{Int}\mskip1.5mu\}\mskip1.5mu\}\mskip1.5mu\}{}\<[E]%
\\
\>[B]{}\Conid{R}\mathrel{=}\{\mskip1.5mu \mskip1.5mu\}{}\<[E]%
\ColumnHook
\end{hscode}\resethooks
\end{minipage}

\caption{Inferring the class context of \ensuremath{\Conid{Stream}} and constructing the substitution for its recursive use}
\label{recursive_stream}
\end{figure}

\subsection{Customized Context Generation}
There are many other possible solutions for inferring the class context of a data type. One direction is to bring \ensuremath{\Conid{TcM}} from the GHC compiler into Template Haskell APIs \cite{Online:haskelldarkart} and thereby reuse GHC's type inference machinery to solve this problem. Roughly speaking, this approach uses \ensuremath{\Varid{unsafeCoerce}} to coerce \ensuremath{\Conid{TcM}} to \ensuremath{\Conid{Q}} so that type inference actions can run within the Template Haskell \ensuremath{\Conid{Q}} monad. We leave this approach to future work.

\subsection{Breaks of the Generation Process} \label{generation_break}
For the generation algorithm, there are special cases in which programmers need to provide a list of type names that should stop the process. The three kinds of APIs also differ slightly from one another. We briefly discuss these issues in this section.

\subsubsection{Deriving of Primitive Types}
In Haskell, we can neither use nor meaningfully want standalone deriving declarations for primitive types such as \ensuremath{\Conid{Int}} and \ensuremath{\Conid{Double}}, since they are defined in terms of unboxed types such as \ensuremath{\Conid{Int}\mathbin{\#}} and \ensuremath{\Conid{Double}\mathbin{\#}}. We also do not need to write empty instance declarations for them, because current generic programming techniques do not apply to primitive types.

However, for the \ensuremath{\Varid{deriving\char95 th}} API, the instance generation function of type \ensuremath{\Conid{Name}\to \Conid{Q}\;[\mskip1.5mu \Conid{Dec}\mskip1.5mu]} may work perfectly well on primitive types. Our \ensuremath{\Varid{mkArity}} example illustrates this.

\subsubsection{Specificity of the \ensuremath{\Conid{Generic}} Class}
\ensuremath{\Conid{Generic}} differs from classes such as \ensuremath{\Conid{Show}} and \ensuremath{\Conid{Eq}} because it does not require the fields of a type to be \ensuremath{\Conid{Generic}} instances. For a data type such as \ensuremath{\mathbf{data}\;\Conid{Blah}\mathrel{=}\Conid{B}\;\Conid{Ratio}\;\Conid{Seq}}, the data constructors of \ensuremath{\Conid{Ratio}} and \ensuremath{\Conid{Seq}} are not exported from their defining modules, so we cannot and do not need to make them instances of \ensuremath{\Conid{Generic}}. The same issue arises for types such as \ensuremath{\Conid{Integer}}, \ensuremath{\Conid{Set}}, and \ensuremath{\Conid{Map}}. This case is common enough that the generation process needs a way to stop on such types. More generally, users may simply not want some types to become \ensuremath{\Conid{Generic}} instances, so the APIs must allow them to specify this choice.

We can summarize the design in two points:
\begin{enumerate}
\item The \ensuremath{\Varid{other\char95 cond}} for \ensuremath{\Varid{deriving\char95 }} and \ensuremath{\Varid{instance\char95 }} should break on primitive types. 
\item A list of type names can be passed to the APIs to break the generation process.
\end{enumerate}

For the first point, it is straightforward to define a function \ensuremath{\Varid{isPrimitive}\mathbin{::}\Conid{Name}\to \Conid{Q}\;\Conid{Bool}} that checks whether a type is primitive and to add it to the break condition for \ensuremath{\Varid{deriving\char95 with}} and \ensuremath{\Varid{instance\char95 with}}. This function is provided by the \ensuremath{\Varid{primitive}} package.

\begin{hscode}\SaveRestoreHook
\column{B}{@{}>{\hspre}l<{\hspost}@{}}%
\column{9}{@{}>{\hspre}l<{\hspost}@{}}%
\column{E}{@{}>{\hspre}l<{\hspost}@{}}%
\>[B]{}\Varid{general\char95 derive}\mathbin{::}\Conid{ClassName}\to \Conid{TypeName}\to [\mskip1.5mu \Conid{TypeName}\mskip1.5mu]\to \Conid{StateT}\;[\mskip1.5mu \Conid{Type}\mskip1.5mu]\;\Conid{Q}\;[\mskip1.5mu \Conid{Dec}\mskip1.5mu]{}\<[E]%
\\
\>[B]{}\Varid{general\char95 derive}\;\Varid{cn}\;\Varid{tn}\;\Varid{bs}\mathrel{=}\mathbf{do}{}\<[E]%
\\
\>[B]{}\hsindent{9}{}\<[9]%
\>[9]{}\mathbin{...}{}\<[E]%
\\
\>[B]{}\hsindent{9}{}\<[9]%
\>[9]{}\Varid{isPrim}\leftarrow \Varid{isPrimitive}\;\Varid{tn}{}\<[E]%
\\
\>[B]{}\hsindent{9}{}\<[9]%
\>[9]{}\mathbf{if}\;\Varid{isInstance}\;\Varid{cn}\;\Varid{t}\mathrel{\vee}\Varid{elem}\;\Varid{t}\;\Varid{ts}\mathrel{\vee}\Varid{isPrim}\mathrel{\vee}\Varid{elem}\;\Varid{tn}\;\Varid{bs}{}\<[E]%
\\
\>[B]{}\hsindent{9}{}\<[9]%
\>[9]{}\mathbin{...}{}\<[E]%
\ColumnHook
\end{hscode}\resethooks

As discussed above, \ensuremath{\Varid{isPrim}} is not needed by \ensuremath{\Varid{deriving\char95 th}}, so we either parameterize this condition or define another \ensuremath{\Varid{general\char95 derive}} for that API.

We can therefore redefine the three kinds of APIs with an additional parameter of type \ensuremath{[\mskip1.5mu \Conid{TypeName}\mskip1.5mu]} so that the process can stop at arbitrary types.

\begin{hscode}\SaveRestoreHook
\column{B}{@{}>{\hspre}l<{\hspost}@{}}%
\column{E}{@{}>{\hspre}l<{\hspost}@{}}%
\>[B]{}\Varid{deriving\char95 with}\mathbin{::}\Conid{ClassName}\to \Conid{TypeName}\to [\mskip1.5mu \Conid{TypeName}\mskip1.5mu]\to \Conid{ContextGenerator}\to \Conid{Q}\;[\mskip1.5mu \Conid{Dec}\mskip1.5mu]{}\<[E]%
\\[\blanklineskip]%
\>[B]{}\Varid{instance\char95 with}\mathbin{::}\Conid{ClassName}\to \Conid{TypeName}\to [\mskip1.5mu \Conid{TypeName}\mskip1.5mu]\to \Conid{ContextGenerator}\to \Conid{Q}\;[\mskip1.5mu \Conid{Dec}\mskip1.5mu]{}\<[E]%
\\[\blanklineskip]%
\>[B]{}\Varid{deriving\char95 th\char95 with}\mathbin{::}(\Conid{ClassName},\Conid{Name}\to \Conid{Q}\;[\mskip1.5mu \Conid{Dec}\mskip1.5mu])\to \Conid{TypeName}\to [\mskip1.5mu \Conid{TypeName}\mskip1.5mu]\to \Conid{Q}\;[\mskip1.5mu \Conid{Dec}\mskip1.5mu]{}\<[E]%
\ColumnHook
\end{hscode}\resethooks

\subsection{Debugging the Generation Process}
In some situations, it can be difficult to determine which types should appear in the break list. At the same time, the resulting error messages may be unhelpful because of Haskell's semantics of imprecise exceptions\cite{DBLP:conf/pldi/JonesRHHM99}. Programmers may also want access to intermediate information produced during generation.

For example, if we need to derive \ensuremath{\Conid{Generic}} for \ensuremath{\Conid{HsModule}} in the \ensuremath{\Varid{ghc}} library, we must exclude 15 types. Finding all of them can be difficult when the programmer is not familiar with the surrounding type declarations.

\begin{hscode}\SaveRestoreHook
\column{B}{@{}>{\hspre}l<{\hspost}@{}}%
\column{8}{@{}>{\hspre}l<{\hspost}@{}}%
\column{9}{@{}>{\hspre}l<{\hspost}@{}}%
\column{28}{@{}>{\hspre}l<{\hspost}@{}}%
\column{42}{@{}>{\hspre}l<{\hspost}@{}}%
\column{57}{@{}>{\hspre}l<{\hspost}@{}}%
\column{72}{@{}>{\hspre}l<{\hspost}@{}}%
\column{86}{@{}>{\hspre}l<{\hspost}@{}}%
\column{E}{@{}>{\hspre}l<{\hspost}@{}}%
\>[B]{}\Varid{deriving\char95 with}\;\,$\textquotesingle\textquotesingle$\!\!\;\Conid{Generic}\;\,$\textquotesingle\textquotesingle$\!\!\;\Conid{HsModule}\;{}\<[E]%
\\
\>[B]{}\hsindent{8}{}\<[8]%
\>[8]{}[\mskip1.5mu \,$\textquotesingle\textquotesingle$\!\!\;\Conid{ShortByteString},\,$\textquotesingle\textquotesingle$\!\!\;\Conid{ForeignPtr},\,$\textquotesingle\textquotesingle$\!\!\;\Conid{Array},{}\<[57]%
\>[57]{}\,$\textquotesingle\textquotesingle$\!\!\;\Conid{TyCon},{}\<[72]%
\>[72]{}\,$\textquotesingle\textquotesingle$\!\!\;\Conid{IORef},{}\<[86]%
\>[86]{}\,$\textquotesingle\textquotesingle$\!\!\;\Conid{Var},{}\<[E]%
\\
\>[8]{}\hsindent{1}{}\<[9]%
\>[9]{}\,$\textquotesingle\textquotesingle$\!\!\;\Conid{STRef},{}\<[28]%
\>[28]{}\,$\textquotesingle\textquotesingle$\!\!\;\Conid{Bag},{}\<[42]%
\>[42]{}\,$\textquotesingle\textquotesingle$\!\!\;\Conid{RealSrcSpan},\,$\textquotesingle\textquotesingle$\!\!\;\Conid{FastZString},\,$\textquotesingle\textquotesingle$\!\!\;\Conid{ByteString},\,$\textquotesingle\textquotesingle$\!\!\;\Conid{Name},{}\<[E]%
\\
\>[8]{}\hsindent{1}{}\<[9]%
\>[9]{}\,$\textquotesingle\textquotesingle$\!\!\;\Conid{UniqDSet},{}\<[28]%
\>[28]{}\,$\textquotesingle\textquotesingle$\!\!\;\Conid{Unique},{}\<[42]%
\>[42]{}\,$\textquotesingle\textquotesingle$\!\!\;\Conid{OccName}\mskip1.5mu]\;\Varid{genInferredContext}{}\<[E]%
\ColumnHook
\end{hscode}\resethooks

For the first issue, we can use functions in \ensuremath{\Conid{\Conid{Debug}.Trace}} to emit more informative error messages. For the second, we can provide another set of APIs that print related information during the generation process. We can add the \ensuremath{\Varid{\char95 debug}} suffix to the three kinds of APIs and define \ensuremath{\Varid{deriving\char95 with\char95 debug}}, \ensuremath{\Varid{instance\char95 with\char95 debug}}, and \ensuremath{\Varid{deriving\char95 th\char95 with\char95 debug}}.

\section{Deriving Multiple Class Instances} \label{derive_multiple_class_instances}

\subsection{Deriving Instances Along Type Classes}
\ensuremath{\Conid{MonadIO}} is a type class for monads that can perform \ensuremath{\Conid{IO}} computations.
Its class hierarchy is shown in \autoref{mondic}:
\begin{figure}
\begin{center}
\ensuremath{\Conid{Functor}\Rightarrow \Conid{Applicative}\Rightarrow \Conid{Monad}\Rightarrow \Conid{MonadIO}}
\end{center}
	\centering
	\theverbbox
	\caption{Monadic class hierarchy}
	\label{mondic}
\end{figure}

If an \ensuremath{\Conid{IO}} monad is wrapped in a \textit{newtype} \ensuremath{\Conid{MIO}} and we want to derive \ensuremath{\Conid{MonadIO}}, all of its superclasses must appear in the deriving clause, as follows:

\begin{hscode}\SaveRestoreHook
\column{B}{@{}>{\hspre}l<{\hspost}@{}}%
\column{E}{@{}>{\hspre}l<{\hspost}@{}}%
\>[B]{}\mathbf{newtype}\;\Conid{MIO}\;\Varid{a}\mathrel{=}\Conid{MIO}\;(\Conid{IO}\;\Varid{a})\;\mathbf{deriving}\;\mathbf{newtype}\;(\Conid{Functor},\Conid{Applicative},\Conid{Monad},\Conid{MonadIO}){}\<[E]%
\ColumnHook
\end{hscode}\resethooks

A more complex example is the derivation of numeric type classes. The hierarchy of numeric classes in Haskell is carefully designed and is shown in \autoref{RealFloatFig}.

\begin{figure}
\begin{hscode}\SaveRestoreHook
\column{B}{@{}>{\hspre}l<{\hspost}@{}}%
\column{E}{@{}>{\hspre}l<{\hspost}@{}}%
\>[B]{}(\Conid{RealFrac}\;\Varid{a},\Conid{Floating}\;\Varid{a})\Rightarrow \Conid{RealFloat}\;\Varid{a}{}\<[E]%
\\
\>[B]{}(\Conid{Real}\;\Varid{a},\Conid{Fractional}\;\Varid{a})\Rightarrow \Conid{RealFrac}\;\Varid{a}{}\<[E]%
\\
\>[B]{}\Conid{Fractional}\;\Varid{a}\Rightarrow \Conid{Floating}\;\Varid{a}{}\<[E]%
\\
\>[B]{}(\Conid{Num}\;\Varid{a},\Conid{Ord}\;\Varid{a})\Rightarrow \Conid{Real}\;\Varid{a}{}\<[E]%
\\
\>[B]{}\Conid{Num}\;\Varid{a}\Rightarrow \Conid{Fractional}\;\Varid{a}{}\<[E]%
\\
\>[B]{}\Conid{Eq}\;\Varid{a}\Rightarrow \Conid{Ord}\;\Varid{a}{}\<[E]%
\ColumnHook
\end{hscode}\resethooks
	\centering
	\theverbbox
	\caption{Numerical type class hierarchy}
	\label{RealFloatFig}
\end{figure}

If one wraps \textit{Float} or \textit{Double} in a \textit{newtype} and wants to derive a \textit{RealFloat} instance, several additional type classes must also be derived, as the following example shows:

\begin{hscode}\SaveRestoreHook
\column{B}{@{}>{\hspre}l<{\hspost}@{}}%
\column{E}{@{}>{\hspre}l<{\hspost}@{}}%
\>[B]{}\mathbf{newtype}\;\Conid{F32}\mathrel{=}\Conid{F32}\;\Conid{Float}\;\mathbf{deriving}\;\mathbf{newtype}\;(\Conid{Floating},\Conid{Fractional},\Conid{Eq},\Conid{Ord},\Conid{Num},\Conid{Real},\Conid{RealFrac},\Conid{RealFloat}){}\<[E]%
\ColumnHook
\end{hscode}\resethooks

\begin{figure}
\begin{hscode}\SaveRestoreHook
\column{B}{@{}>{\hspre}l<{\hspost}@{}}%
\column{9}{@{}>{\hspre}l<{\hspost}@{}}%
\column{17}{@{}>{\hspre}l<{\hspost}@{}}%
\column{25}{@{}>{\hspre}l<{\hspost}@{}}%
\column{E}{@{}>{\hspre}l<{\hspost}@{}}%
\>[B]{}\Varid{derive\char95 superclass}\mathbin{::}\Conid{Maybe}\;\Conid{DerivStrategy}\to \Conid{ClassName}\to \Conid{TypeName}\to \Conid{StateT}\;[\mskip1.5mu \Conid{Type}\mskip1.5mu]\;\Conid{Q}\;[\mskip1.5mu \Conid{Dec}\mskip1.5mu]{}\<[E]%
\\
\>[B]{}\Varid{derive\char95 superclass}\;\Varid{s}\;\Varid{cn}\;\Varid{tn}\;\Varid{bs}\mathrel{=}\mathbf{do}{}\<[E]%
\\
\>[B]{}\hsindent{9}{}\<[9]%
\>[9]{}\Varid{ts}\leftarrow \Varid{get}{}\<[E]%
\\
\>[B]{}\hsindent{9}{}\<[9]%
\>[9]{}\mathbf{let}\;\Varid{t}\mathrel{=}\Varid{constructType}\;\Varid{tn}{}\<[E]%
\\
\>[B]{}\hsindent{9}{}\<[9]%
\>[9]{}\mathbf{if}\;\Varid{isInstance}\;\Varid{cn}\;\Varid{t}\mathrel{\vee}\Varid{elem}\;\Varid{t}\;\Varid{ts}{}\<[E]%
\\
\>[9]{}\hsindent{8}{}\<[17]%
\>[17]{}\mathbf{then}\;\Varid{return}\;[\mskip1.5mu \mskip1.5mu]{}\<[E]%
\\
\>[9]{}\hsindent{8}{}\<[17]%
\>[17]{}\mathbf{else}\;\mathbf{do}{}\<[E]%
\\
\>[17]{}\hsindent{8}{}\<[25]%
\>[25]{}\Varid{cxt}\leftarrow \Varid{genClassContext}\;\Varid{cn}\;\Varid{tn}{}\<[E]%
\\
\>[17]{}\hsindent{8}{}\<[25]%
\>[25]{}\mathbf{let}\;\Varid{dec}\mathrel{=}\Varid{genDecl}\;\Varid{s}\;\Varid{cn}\;\Varid{cxt}\;\Varid{t}{}\<[E]%
\\
\>[17]{}\hsindent{8}{}\<[25]%
\>[25]{}\Varid{modify}\;(\Varid{t}\mathbin{:}){}\<[E]%
\\
\>[17]{}\hsindent{8}{}\<[25]%
\>[25]{}\Varid{names}\leftarrow \Varid{lift}\mathbin{\$}\Varid{getSuperClassNames}\;\Varid{cn}{}\<[E]%
\\
\>[17]{}\hsindent{8}{}\<[25]%
\>[25]{}\Varid{decs}\leftarrow \Varid{mapM}\;(\Varid{general\char95 superclass}\;\Varid{cn})\;\Varid{ns}{}\<[E]%
\\
\>[17]{}\hsindent{8}{}\<[25]%
\>[25]{}\Varid{return}\mathbin{\$}\Varid{dec}\mathbin{:}\Varid{concat}\;\Varid{decs}{}\<[E]%
\\[\blanklineskip]%
\>[B]{}\Varid{getSuperClassNames}\mathbin{::}\Conid{Name}\to \Conid{Q}\;[\mskip1.5mu \Conid{Name}\mskip1.5mu]{}\<[E]%
\ColumnHook
\end{hscode}\resethooks
\caption{Algorithm for deriving instances along superclasses}
\label{superclass_alg}
\end{figure}
Apparently, this process can be automated with a method similar to the one above. The algorithm is almost the same as the one presented in the previous section. It should be implemented as a meta-programming function with three parameters and result type \ensuremath{\Conid{Q}\;[\mskip1.5mu \Conid{Dec}\mskip1.5mu]}: an optional deriving strategy\cite{deriving_strategies}, a type name, and a class name. The process is listed in Figure \ref{superclass_alg}. \ensuremath{\Varid{getSuperClassNames}} is a meta function that reifies a class name and returns its superclasses, possibly as an empty list. For classes such as \ensuremath{\Conid{MonadIO}} and \ensuremath{\Conid{RealFloat}}, the deriving strategy is normally \textit{newtype}. For classes such as \ensuremath{\Conid{Ord}}, by contrast, the default \textit{stock} strategy, or no explicit strategy at all, should be used. Accordingly, \ensuremath{\Varid{derive\char95 superclass}} can be used to define the following two APIs:

\begin{hscode}\SaveRestoreHook
\column{B}{@{}>{\hspre}l<{\hspost}@{}}%
\column{32}{@{}>{\hspre}l<{\hspost}@{}}%
\column{E}{@{}>{\hspre}l<{\hspost}@{}}%
\>[B]{}\Varid{strategy\char95 deriving\char95 superclasses}\mathbin{::}\Conid{DerivStrategy}\to \Conid{ClassName}\to \Conid{TypeName}\to \Conid{Q}\;[\mskip1.5mu \Conid{Dec}\mskip1.5mu]{}\<[E]%
\\
\>[B]{}\Varid{deriving\char95 superclasses}{}\<[32]%
\>[32]{}\mathbin{::}\Conid{ClassName}\to \Conid{TypeName}\to \Conid{Q}\;[\mskip1.5mu \Conid{Dec}\mskip1.5mu]{}\<[E]%
\ColumnHook
\end{hscode}\resethooks

\section{Future Work}
This work extends Haskell's deriving mechanism to support composite types and classes with many superclasses, but there is still substantial room for improvement. We briefly list several directions for future work.
\begin{itemize}
\item Although it would be straightforward to define a function that derives class instances along both composite types and superclasses, we do not currently consider this capability essential, and it can be implemented later.
\item When deriving \ensuremath{\Conid{Generic}} instances, the process might automatically stop at data constructors of types whose constructors are not exported, without requiring programmers to list those types explicitly.
\item The list of type names that break the generation process could be replaced with a predicate of type \ensuremath{\Conid{Name}\to \Conid{Bool}} for greater flexibility, or even \ensuremath{\Conid{Name}\to \Conid{Q}\;\Conid{Bool}}, allowing predicates such as ``derive only types in the current module'' or ``derive only types in modules \ensuremath{\Conid{\Conid{GHC}.A}} and \ensuremath{\Conid{\Conid{GHC}.B}}''.
\item For simplicity, this work considers only classes of kind ${\star \rightarrow Constraint}$ and higher-order type classes of kind ${(\star \rightarrow \star) \rightarrow Constraint}$. Deriving instances for multi-parameter type classes remains open for future exploration.
\item We applied this technique to the AST in the \ensuremath{\Varid{ghc}} package and found that the relevant data types are spread across about 30 modules. Importing each of them can itself become tedious, so a top-down import mechanism may also be worth exploring.
\end{itemize}

\section{Conclusion}
In this work, we propose a top-down deriving mechanism that can derive multiple class instances along composite data types or class hierarchies. We present algorithms for generating instance declarations and discuss several methods for generating class contexts. Although our work focuses on Haskell, we believe that its core ideas can be transferred to languages beyond Haskell.

\begin{acks}
Special thanks to Ryan Scott for resourceful ideas that helped solve the problems we encountered. We also thank Richard Eisenberg for timely help with issues that arose during the implementation of this work. We are grateful to Tesla Zhang, Zirun Zhu for discussing these ideas and offering valuable feedback. Clerk Ma also helped with the typesetting of this work.
\end{acks}

\bibliographystyle{ACM-Reference-Format}
\bibliography{sample-base}
\appendix

\end{document}